\documentstyle[12pt]{article}

\begin{document}
\thispagestyle{empty}
\begin{center}
\LARGE \tt \bf{Supersymmetric Potentials in Inflationary Einstein-Cartan-Brans-Dicke Cosmology}
\end{center}
\vspace{1cm}
\begin{center} {\large L.C. Garcia de Andrade\footnote{Departamento de Fisica Teorica-IF-UERJ-CEP:20550-013,Rio de Janeiro,RJ}}
\end{center}
\vspace{1.0cm}
\begin{abstract}
Cosmological inflation is discussed in the realm of Einstein-Cartan-Brans-Dicke (ECBD) gravity by constructing an effective inflaton potential and computing the number of e-folds.It is shown that spin-torsion density contributes to a decrease on the number of e-folds of inflation when the ratio of spin-torsion density to matter density is aprecciable as happens in the Early Universe.Quantum fluctuations of spin and inflatons are also investigated.Our results seems to be in agreement with Palle's proposal that quantum fluctuations of spin may trigger the primordial era of the universe.The main esult of the paper is the appearence of a supersymmetric type inflaton potential from the ECBD gravity where the coefficient of the one-loop type correction is a constant spin-torsion density.  
\end{abstract}      
\vspace{1.0cm}       
\begin{center}
\Large{PACS number(s) : 0420,0450,98.80.Cq}
\end{center}
\newpage
\pagestyle{myheadings}
\markright{\underline{Supersymmetric Inflaton potentials}}
\paragraph*{}
\section{Introduction}
Recently S.Kim \cite{1} proposed a cosmological model endowed with torsion, scalar fields and what he called an inflationary ECBD cosmology.Nevertheless in the most modern and common sense inflation does need to an inflaton field and potential to trigger inflation \cite{2,3}.However Kim ECBD inflation does not possess such a feature and he considers just vacuum inflation by making use of the condition in the fluid $p+{\rho}=0$ where p and ${\rho}$ are respectively the pressure and matter density of the spinning fluid \cite{4}.In this Letter we propose to remedy this fact by constructing an effective inflaton potential in two distinct cases.In the first section we address a dust composed of spins where the matter density and the spin-torsion density tensor do not depend on the inflaton field.In this case we compute the number of e-folds of inflation and show that the presence of spin-torsion induces a decrease on the e-folds number when the ratio of spin-torsion density to the matter density is appreciable which happens in the early universe where inflation occurs.In section 2 the radiation filled spinning fluid universe is considered where the inflaton fluctuation equation has the spin-torsion density as a source and in this way can be considered to trigger quantum fluctuations of the inflaton field.
\section{Effective Inflaton Potential in ECBD}
We start this section by reproducing the Lagrangean and field equations of Kim's theory as
\begin{equation} 
L=\int{d^{4}x\sqrt{-g}(-{\phi}R+{\omega}\frac{{\partial}^{\mu}{\phi}{\partial}_{\mu}{\phi}}{\phi})}
\label{1}
\end{equation}
where ${\mu}=0,1,2,3$.Note that the only way to couple scalar fields to torsion is through non-minimal coupling \cite{5}.The field equations are given by
\begin{equation}
H^{2}=\frac{8{\pi}}{3{\phi}}({\rho}-\frac{2{\pi}{\sigma}^{2}}{\phi})+\frac{1}{6}({\omega}+\frac{3}{2})\frac{{\dot{\phi}}^{2}}{{\phi}^{2}}+H\frac{{\dot{\phi}}}{\phi}
\label{2}
\end{equation}
\begin{equation}
\frac{\ddot{R}}{R}=-\frac{4{\pi}}{3{\phi}}({\rho}+3p-\frac{8{\pi}{\sigma}^{2}}{\phi})-\frac{1}{3}({\omega}+\frac{3}{2})\frac{{\dot{\phi}}^{2}}{{\phi}^{2}}+\frac{{\ddot{\phi}}}{2{\phi}}
\label{3}
\end{equation}
\begin{equation}
\ddot{\phi}+3H\dot{\phi}=\frac{4{\pi}}{{\omega}}({\rho}-3p-\frac{8{\pi}{\sigma}^{2}}{\phi})
\label{4}
\end{equation}
where ${\phi}$ is the inflaton field and $H=\frac{\dot{R}}{R}$ is the Hubble expansion while R is the cosmic scale factor.Units are used here where  the gravitational constant $G=1$.Conservation laws for matter and spin are taken separetly as
\begin{equation}
\dot{{\rho}}=-3H({\rho}+p)
\label{5}
\end{equation}
\begin{equation}
\dot{{\sigma}^{2}}=-6H{\sigma}^{2}
\label{6}
\end{equation}
The effective inflaton potential $V_{eff}({\phi})$ can be easily constructed can be easily considered by comparison between equation (\ref{4}) and the inflaton equation
\begin{equation}
\ddot{\phi}+3H\dot{\phi}=-\frac{{\partial}V_{eff}}{{\partial}{\phi}}
\label{7}
\end{equation}
which yields
\begin{equation}
\frac{{\partial}V_{eff}}{{\partial}{\phi}}=-\frac{4{\pi}}{{\omega}}({\rho}-3p-\frac{8{\pi}{\sigma}^{2}}{\phi})
\label{8}
\end{equation}
A simple solution of this equation to determine the inflaton potential can be given in the case the spin and matter densities are constants.In this case we obtain the following inflaton potential
\begin{equation}
V_{eff}=-\frac{4{\pi}}{\omega}({\rho}_{0}-3p_{0}){\phi}+\frac{32{\pi}}{\omega}{{\sigma}_{0}}^{2}ln{\phi}
\label{9}
\end{equation}
This inflaton potential is similar to the Colemann-Weinberg potential and in fact agrees with hybrid inflation supersymmetric one-loop correction potential.This result was to a certain extent already expected since the bosonic (inflaton fields) and Fermionic (spinning fluid) sectors are present in the theory.In fact recently another example of supersymmetric behaviour on Einstein-Cartan gravity has appeard in the form of supersymmetric domain walls \cite{6}.Computation of the e-folds number of inflation
\begin{equation}
N({\phi})={M^{-2}}_{Pl}\int{\frac{V}{V'}d{\phi}}
\label{10}
\end{equation}
yields
\begin{equation} 
N({\phi})={M^{-2}}_{Pl}[\frac{\phi}{2}+4{\pi}{\omega}{\phi}(ln{\phi}-1)]
\label{11}
\end{equation}
Here $M_{Pl}$ is the Planck mass.To obtain expresion (\ref{11}) we made the approximation
\begin{equation}
\frac{4{\pi}}{\omega}<<{{\sigma}_{0}}^{2}
\label{12}
\end{equation}
Note that since general relativity tests impose a constraint on the Brans-Dicke inflation as ${\omega}>>>500$ expression (\ref{12}) also implies a constraint on the spin-torsion density that is well within the values of torsion found in the Early Universe where inflation occurs.Let us now examine the de Sitter inflation where $H=H_{0}=constant$.In this case the field equations yield after a long but straightforward computation the equation for the inflaton fluctuations
\begin{equation}
{\delta}\ddot{\phi}+3[3{H_{0}}^{2}-4{\pi}{\rho}_{0}]{\delta}{\phi}=\frac{8{\pi}}{3}{\delta}{\rho}+\frac{16{\pi}^{2}}{{\phi}_{0}}{\delta}{{\sigma}^{2}}
\label{13}
\end{equation}
where we have assumed that the inflaton matter and spin-torsion density fluctuation are given by
\begin{equation}
{\phi}={\phi}_{0}+{\delta}{\phi}
\label{14}
\end{equation}
\begin{equation}
{\rho}={\rho}_{0}+{\delta}{\rho}
\label{15}
\end{equation}
\begin{equation}
{\sigma}^{2}={{\sigma}_{0}}^{2}+{\delta}{{\sigma}}^{2}
\label{16}
\end{equation}
Note that in the case ${\delta}{\rho}={\delta}{\sigma}^{2}=0$ yields
\begin{equation}
{\delta}\ddot{\phi}+3[3{H_{0}}^{2}-4{\pi}{\rho}_{0}]{\delta}{\phi}=0
\label{17}
\end{equation}
which yields the following solution
\begin{equation}
{\delta}{\phi}=Acos{\alpha}t+Bsin{\alpha}t
\label{18}
\end{equation}
where A and B are integrationm constants and 
\begin{equation}
{\alpha}=3[3{H_{0}}^{2}-4{\pi}{\rho}_{0}]
\label{19}
\end{equation}
which is similar to what happens well before the horizon exit.In fact we see here that the inflaton fluctuation indeed oscillates.Another interesting case is when the inflaton do not fluctuates or ${\delta}{\phi}=0$.In this case the inflaton equation reduces to
\begin{equation}
{\delta}{\rho}=-2{\pi}\frac{{\delta}{\sigma}^{2}}{{\rho}_{0}}
\label{20}
\end{equation}
Since from the spin and matter conservation laws above one obtains for de Sitter inflation
\begin{equation}
{\delta}{\rho}=e^{{\beta}t}
\label{21}
\end{equation}
\begin{equation}
{\delta}{\sigma}^{2}=e^{{\beta}t}
\label{22}
\end{equation}
where ${\beta}=-3H_{0}$.Substitution of these values on equation (\ref{16}) yields
\begin{equation}
{\delta}{\phi}{\alpha} t
\label{23}
\end{equation}
This expression shows that the inflaton is unstable.
\section{The Radiation era in ECBD Inflation}
In the radiation era of the universe some interesting but expected phenomenon happens since in the radiation era the role of spin-torsion is very strong we shall find that the spin-torsion is a source of inflaton fluctuation but also depends on the inflaton potential.To this end let us consider that in the radiation era the presure is given by $p=\frac{1}{3}{\rho}$.Substitution of this expression into the field equations yield
\begin{equation}
H_{0}^{2}=\frac{32{\pi}}{3{\phi}^{2}}{\sigma}^{2}-\frac{1}{3}({\omega}+\frac{3}{2})\frac{{\dot{\phi}}^{2}}{{\phi}^{2}}+\frac{1}{2}\frac{\ddot{\phi}}{\phi}
\label{24}
\end{equation}
\begin{equation}
2H_{0}^{2}=\frac{16{\pi}}{3{\phi}}({\rho}-\frac{2{\pi}}{\phi}{\sigma}^{2})+\frac{1}{3}({\omega}+\frac{3}{2})+2H_{0}\frac{\dot{\phi}}{\phi}
\label{25}
\end{equation}
Summing up these two equations we obtain
\begin{equation}
{\sigma}^{2}({\phi})=\frac{3}{32{\pi}^{2}}{\phi}^{2}
\label{26}
\end{equation}
This allow us now to compute the effective potential above for a inflaton potential where the spin-torsion density depends upon the inflaton field.Substitution of expression (\ref{26}) into (\ref{8}) and after integration yields
\begin{equation}
V_{eff}|_{spin}=\frac{9{\pi}}{16{\omega}}{\phi}^{2}
\label{27}
\end{equation}
which is a simple matter type inflaton potential.The method we have used here is very similar to the reconstruction of the inflaton potential although much more simple.  
\section*{Acknowledgments}
\paragraph*{}
Thanks are due to Prof.J.A.Hehlayel and Prof.I.L.Shapiro , for their interest on the subject of this paper and for helpful discussions.I am very much indebt to CNPq. (Brazilian Government Agency) for financial support.


\begin{thebibliography}{6}
\bibitem{1}Y.S.Kim,Nuovo Cimento B,(1999).
\bibitem{2}D.Palle,Primordial Fluctuations in Einstein-Cartan gravity and COBE data,astro-ph/9811408.
\bibitem{3}D.Lyth and R.Liddle,Cosmological Inflation and Structure Formation (2000).
\bibitem{4}V.de Sabbata and C.Sivaram,Spin and Torsion in Gravitation,(1994)World Scientific. 
\bibitem{5}I.L.Shapiro,On the Physics  of torsion,Physics Reports,to appear.
\bibitem{6}L.C.Garcia de Andrade,Class. and Quantum Gravity,16,(1999),2097 and Phys.Lett.B 28(1999).
\end{thebibliography}
\end{document}